 \documentclass[manuscript]{aastex}

\def\lesssim{\mathrel{\hbox{\rlap{\hbox{\lower4pt\hbox{$\sim$}}}\hbox{$<$}}}}
\def\gtrsim{\mathrel{\hbox{\rlap{\hbox{\lower4pt\hbox{$\sim$}}}\hbox{$>$}}}}

\usepackage{graphicx}
\usepackage{color}
\usepackage{natbib}
\begin{document}

\title{Swift J1644+57: A White Dwarf Tidally Disrupted by a $10^4 M_{\odot}$ Black Hole?}
\shorttitle{A Tidally-Disrupted White Dwarf}

\author{Julian H. Krolik}
\affil{Physics and Astronomy Department\\
Johns Hopkins University\\ 
Baltimore, MD 21218}

\and

\author{Tsvi Piran}
\affil{Racah Institute of Physics\\
Edmond J. Safra Campus\\
Hebrew University of Jerusalem\\ 
Jerusalem 91904, Israel}

\email{jhk@jhu.edu; tsvi@phys.huji.ac.il}

\begin{abstract}
We propose that the remarkable object Swift J1644+57, in which multiple recurring
hard X-ray flares were seen over a span of several days, is a system in which a white dwarf
was tidally disrupted by an intermediate mass black hole.  Disruption of a white
dwarf rather than a main sequence star offers a number of advantages in understanding
the multiple, and short, timescales seen in the light curve of this system.
In particular, the short internal dynamical timescale of a white dwarf offers
a more natural way of understanding the short rise times ($\sim 100$~s) observed.
The relatively long intervals between flares ($\sim 5 \times 10^4$~s) may also
be readily understood as the period between successive pericenter passages of the
remnant white dwarf.  In addition, the expected jet power is larger when a white
dwarf is disrupted.  If this model is correct, the black hole responsible must
have mass $\lesssim 10^5 M_{\odot}$.
\end{abstract}


\section{Introduction}

\begin{figure}
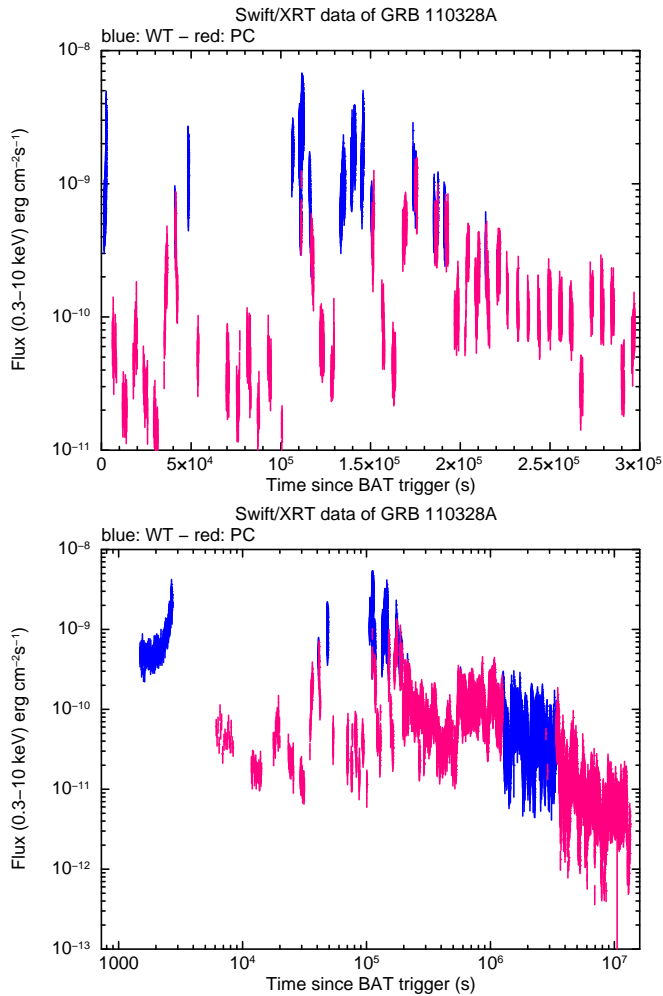


\includegraphics[width=0.4\textwidth,angle=-90]{fig1a.ps}

\includegraphics[width=0.4\textwidth,angle=-90]{fig1b.eps}

\caption{Long-term Swift XRT light curves in the 3--10 keV band (blue: WT, red: PC).
(Top) Linear in time representation of the first 300,000~s, illustrating the
recurring brief flares that gradually widen.  (Bottom)
Logarithmic in time representation of the entire light curve as of 29 August 2011,
five months after activity began.  Both these light curves and those in
the following figure were generated using the graphing tools of the online Swift Data
Repository \citep{Evans07}.
\label{fig:longterm}}
\end{figure}

\begin{figure}
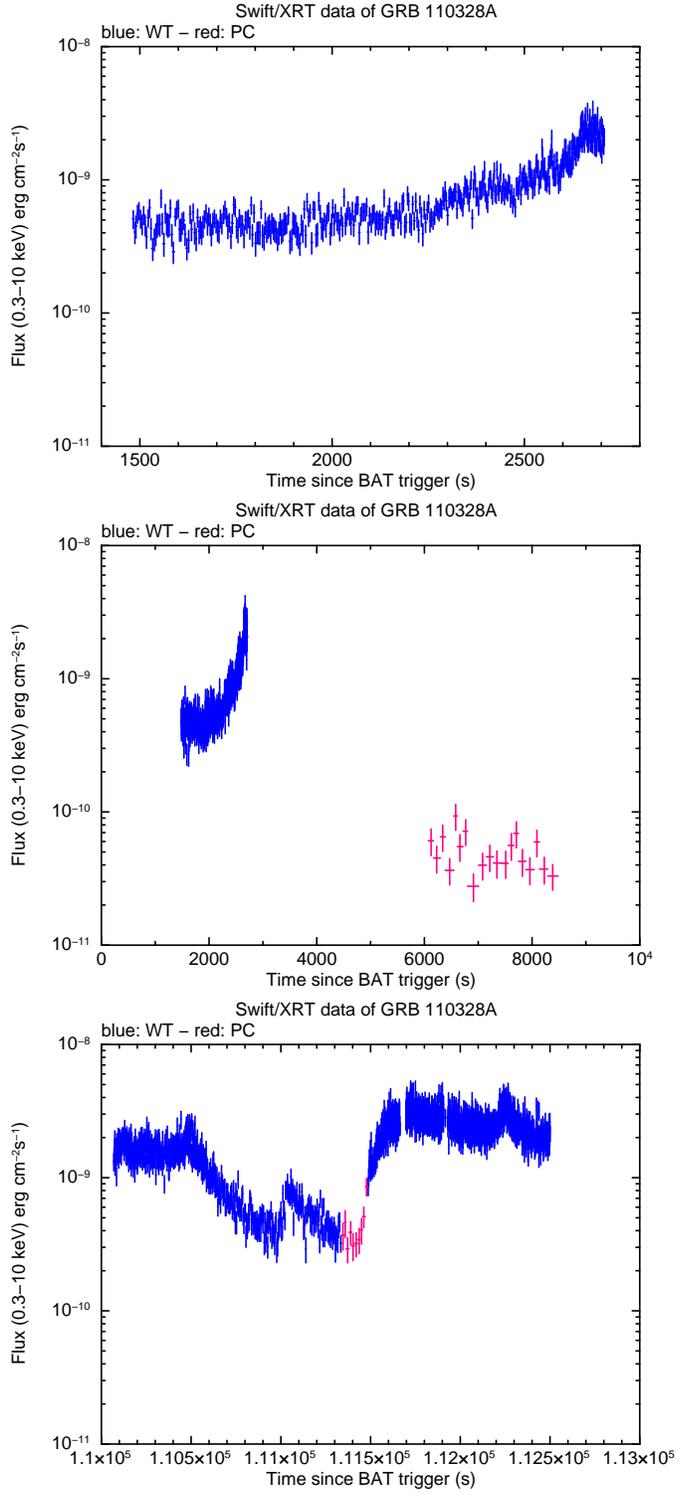

\includegraphics[width=0.4\textwidth,angle=-90]{fig2a.ps}

\includegraphics[width=0.4\textwidth,angle=-90]{fig2b.eps}

\includegraphics[width=0.4\textwidth,angle=-90]{fig2c.ps}

\caption{Detailed time structure of the flares. (Top) Short timescale
structure of the first flare.  (Middle) Overview of the first flare.
(Bottom) Short timescale structure in the flare $\simeq 110,000$~s after the
BAT trigger.  Note that the entire span of the data in the bottom panel is only 1500~s.
\label{fig:shortterm}}
\end{figure}

   On 28 March 2011, the {\it Swift} Burst Alert Telescope (BAT) detected a most
unusual object, Swift J164449.3+573451 \citep{levan11,burrows11}.  Although in
many ways this object (whose name we abbreviate to Swift J1644+57) initially
appeared to resemble a classical $\gamma$-ray burst, its light curve soon showed that
it was quite different.  Still bright more than $6 \times 10^6$~s after the
initial trigger (see Fig.~\ref{fig:longterm}), in its initial activity it
exhibited repeated extremely short timescale flares (see Fig.~\ref{fig:shortterm}).
After holding roughly steady for $\simeq 700$~s, the flare causing the BAT trigger rose a factor
of 10 in flux over the next $\simeq 400$~s.  Less than 1000~s later, the flux had
fallen by a factor of 20, and $\sim 10^4$~s after that, although still detectable,
the flux was only $0.5\%$ of what it had been at the peak.   Most surprisingly, there
was a comparable flare $\simeq 50000$~s later, similarly lasting for only $\sim 1000$~s,
and a third, slightly brighter than the first, $\simeq 60000$~s after that.  Like
the first two, the third flare continued to show very short timescale variation.

Several more brief flares of comparable
brightness followed, likewise separated by periods of flux two orders of magnitude
weaker.  Gradually the duration of the flares stretched and their amplitudes diminished,
until after $\simeq 2 \times 10^5$~s the system began a long, gradual decline in flux
in which it has remained bright enough to be detected for more than
$1.5 \times 10^7$~s.
Thus, there appear to be a number of characteristic
timescales of variation, spread over quite a wide dynamic range: rise-times as short
as $\sim 100$~s; flare durations $\sim 1000$--10000~s; quiescent periods $\sim 5 \times 10^4$~s
long; and a total event duration
of more than $10^7$~s.
{After $10^6$ s the best power-law fit to the light curve is that the observed flux
decays $\propto t^{-4/3}$, but its large variability and spectral changes
mean that it might still be consistent with  the expected $t^{-5/3}$ bolometric decay.}

Still another surprise came when BAT data {\it preceding} the trigger were
examined.  Approximately 3 days earlier, there was a precursor whose peak
flux was $\sim 0.07$ times the flux of the first flare \citep{burrows11}.
Searches for earlier episodes of emission from the same source in archival data
revealed only upper limits. 

   Optical, near-infrared, and radio observations combine to show that the source
lies within 150pc of the center of a galaxy at $z=0.354$ \citep{levan11}.  The peak
luminosity associated with these flares (interpreted as isotropic) is then
$\simeq 4 \times 10^{48}$~erg~s$^{-1}$ \citep{burrows11}.  Even at its faintest
detected level, the (isotropic) luminosity is $\simeq 1 \times 10^{46}$~erg~s$^{-1}$
\citep{burrows11}.  Although relativistic beaming
{that could enhance the flux by a factor $\sim 100$}
is likely \citep{burrows11}, the beaming-corrected power is still
very large: $\sim 4 \times 10^{46}$~erg/s at the peak, and a total emitted energy
$\sim 5 \times 10^{51}$~erg  within the first $1 \times 10^7$~s of the
event if the XRT flux is $\sim 1/3$ of bolometric, as suggested by \cite{bloom11}.

    During flaring episodes, the observed $\nu F_\nu$ is greatest between 10 and 100~keV;
it appears to be strongly absorbed by interstellar gas below $\simeq 5$~keV in the
rest-frame.  By the time the near-IR measurements were performed, 2--4 days after the
event began, their fluxes were $\sim 10^{-4}$ of the hard X-ray flux at its peak, and
$\sim 10^{-2}$ of its flux during quiescent periods \citep{burrows11}.

    Within six weeks of its discovery, several models were proposed to explain it,
most concentrating on a picture in which a main sequence star was tidally
disrupted by passing too close to a $10^6$--$10^7 M_{\odot}$ black hole
\citep{bloom11,burrows11,cannizzo11,socrates11}.  Although these models differ from one
another in some of their details, they share a common outline: A main sequence
star is thoroughly disrupted as it passes near a black hole, and much of its
mass is distributed into an accretion disk around the black hole.  A powerful
jet is then created, whose radiation, created by the synchro-Compton mechanism,
dominates what we observe.  In this
picture, the rise time of the flares is interpreted as indicating (after allowance
for relativistic effects) the size of individual knots in the jet, and these
are related to the gravitational radius of the black hole.  The duration
of the entire event is thought to indicate either the expected $t^{-5/3}$ scaling
\citep{burrows11} due to the orbital period distribution of tidal streams
\citep{rees88,phinney89} or the inflow time of the accretion disk \citep{socrates11}.

     Although plausible in many respects, this consensus model leaves a number of
questions unanswered.  The tidal disruption radius in units of the black hole
gravitational radius is
\begin{equation}\label{eq:rtrg}
R_T/R_g = R_* (k/f)^{1/6}(M_{BH}/M_*)^{1/3}/(GM_{BH}/c^2) = 50 (k/f)^{1/6}
M_{BH,6}^{-2/3} {\cal M}_*^{2/3},
\end{equation}
where $k$ is the apsidal motion constant (determined by the star's radial density
profile) and $f$ is its binding energy in units of $GM_*^2/R_*$ \citep{phinney89}.
Here ${\cal M}_*$ is the mass of the star in solar units, and we have taken the approximate
scaling that $R_* \propto M_*$ on the main sequence.  The ratio $k/f$ is $\simeq 0.02$
for radiative stars, but 0.3 for convective stars \citep{phinney89}.  The circular
orbital period at the tidal radius is then
proportional to ${\cal M}_*$ and {\it independent} of $M_{BH}$ because the dynamical
time at the tidal radius is always $\sim (G\rho_*)^{-1/2}$, and on the main sequence
the mean stellar density $\rho_* \propto M_*^{-2}$:
\begin{equation}
P_{\rm orb}(R_T) = 1.0 \times 10^4 {\cal M}_* \hbox{~s}.
\end{equation}

However, the {\it actual} orbital period of matter captured in tidal disruption is likely
to be considerably longer.  As pointed out by \cite{rees88}, if the disrupted star approached
on a nearly-parabolic orbit, the tidal streams follow highly-eccentric elliptical orbits
with a wide range of energies, and therefore of semi-major axes and orbital periods.  Their mean
energy is likely to be comparable to the star's self-gravitational binding energy because the
gravitational force of the black hole does that much work expanding and disrupting the star.
On the other hand, tidally-induced rotation and the gradient of the gravitational
potential across the star can lead to some gas being trapped on orbits with
semi-major axis as small as $a_d \sim R_T (M_{BH}/M_*)^{1/3}$, even while the
typical stream's orbit is larger by a factor $(M_{BH}/M_*)^{1/3}$.  It follows that even
the shortest orbital period is larger than that of a circular orbit at $R_T$ by
$\sim (M_{BH}/M_*)^{1/2}$, here a factor $\sim 10^3$, and the typical period likely
another factor of $\sim 10^3$ longer than that.  Intersections between stream orbits
could lead to conversion of orbital energy to heat, diminishing these orbital
periods, but in no case would they become shorter than $P_{\rm orb}(R_T)$.
In fact, numerical simulation of the disruption of a main sequence star by a $10^6 M_{\odot}$
black hole \citep{Ayal+00} shows a rather continuous accretion rate with a rise time of a few
times $10^5$~s  and an overall duration of a few times $10^6$~s, as expected from these
analytic estimates. 

It is hard to reconcile these timescales to those seen in the lightcurve.  The rise
time in the consensus model is said to reflect the light-crossing time across the
black hole's horizon, but it is not clear what dynamics link that quantity to
triggering a flare, nor is there any natural explanation for the flare duration.
The circular orbital period at $R_T$ is comparable to the inter-flare interval,
but the orbital period of the tidal streams following eccentric orbits,
which is the timescale at which the $t^{-5/3}$ decay begins, appears to be at
least two orders of magnitude longer than the time at which the flares merged
into a smoother lightcurve.
These difficulties have led some (e.g., \cite{cannizzo11}) to pose special
requirements on this model.  They suggest that the pericenter distance must not
be a great deal larger than the black hole's ISCO, so that inflow is largely
dynamical.  If so, the black hole mass is $\sim 10^7 M_{\odot}$.
By the estimate of Equation~\ref{eq:rtrg}, the pericenter distance must then be a
small fraction of $R_T$.

     Another question is how exactly the accretion drives a jet.  If, as is generally
believed, jets associated with black holes are powered by some variant of the
Blandford-Znajek mechanism \citep{bz77,mck05,hk06}, substantial magnetic field
must be attached to the black hole horizon.  The Poynting luminosity in the jet is then
$\sim \phi B^2 r_g^2 c$, where $\phi$ is a dimensionless quantity that depends on the
field geometry and increases with black hole spin parameter $a/M$, but is generally
significantly less than unity.  In this object, the field on the horizon must then be
$\sim 1 \times 10^6 \phi^{-1/2} L_{45}^{1/2} M_{BH,6}^{-1}$~G.  When the system
has a long lifespan (e.g., in AGN), the magneto-rotational instability stirs MHD turbulence
in the accretion disk and builds the magnetic field; numerical simulations of MRI-driven
MHD turbulence show that generically $\sim 10$ orbital periods are required
to reach saturation \citep{stone96}.  Accretion may also accumulate
magnetic flux on the horizon \citep{beckwith09}; its build-up rate depends, of course,
on the structure of the large-scale magnetic field.
Here, the accretion flow must plunge into the black hole on a dynamical time, and it
is unclear whether such a strong field could be generated.

      Still another problem raised by this model is how to understand the very large
amplitude and very rapidly-varying flares.   As recognized by \cite{bloom11}, relativistic
jets in blazars behave very differently: Their characteristic fluctuation amplitudes
are more typically factors of a few than factors of a few orders of magnitude, and their
duty cycle at high flux is usually considerably greater than it is during the first
several flaring episodes of this event.  One could rephrase this question as, ``What
generates the extremely bright, compact, and short-lived knots that in this model
are assumed to account for the flares?"

\begin{figure}
\includegraphics[width=12cm,angle=90]{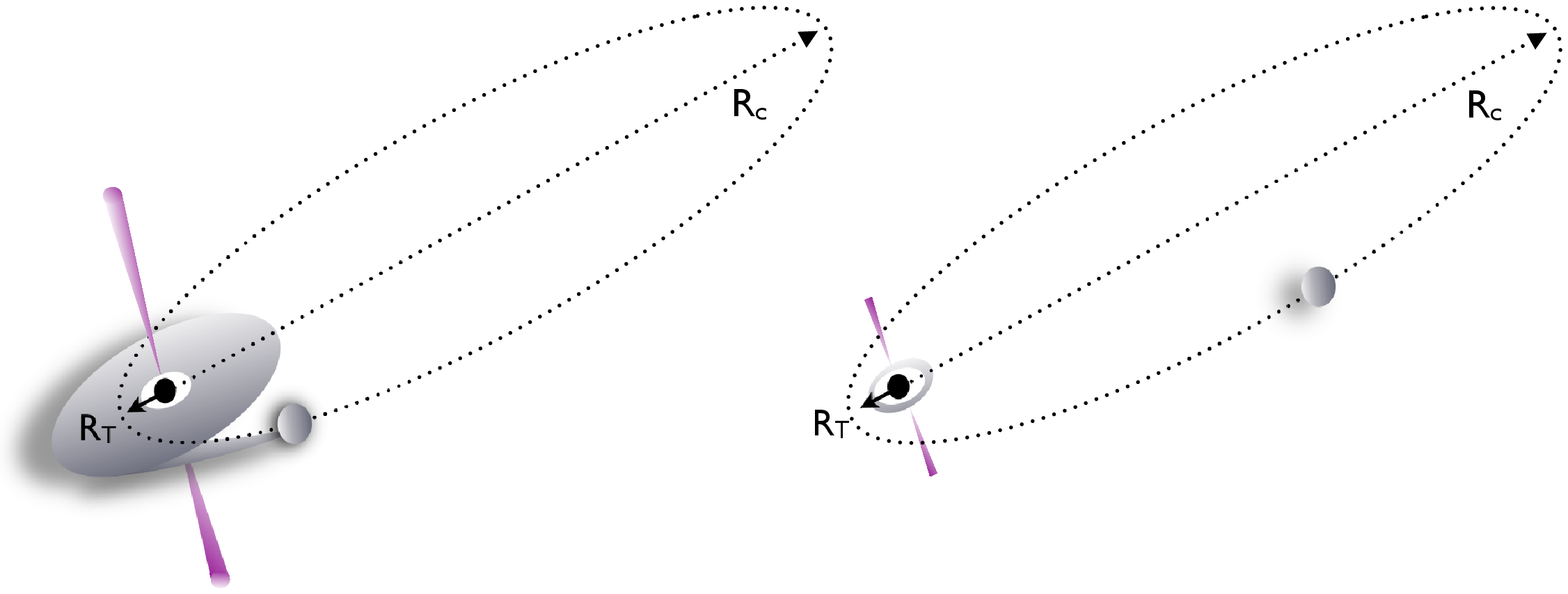}
\vskip -4cm
\includegraphics[width=12cm,angle=90]{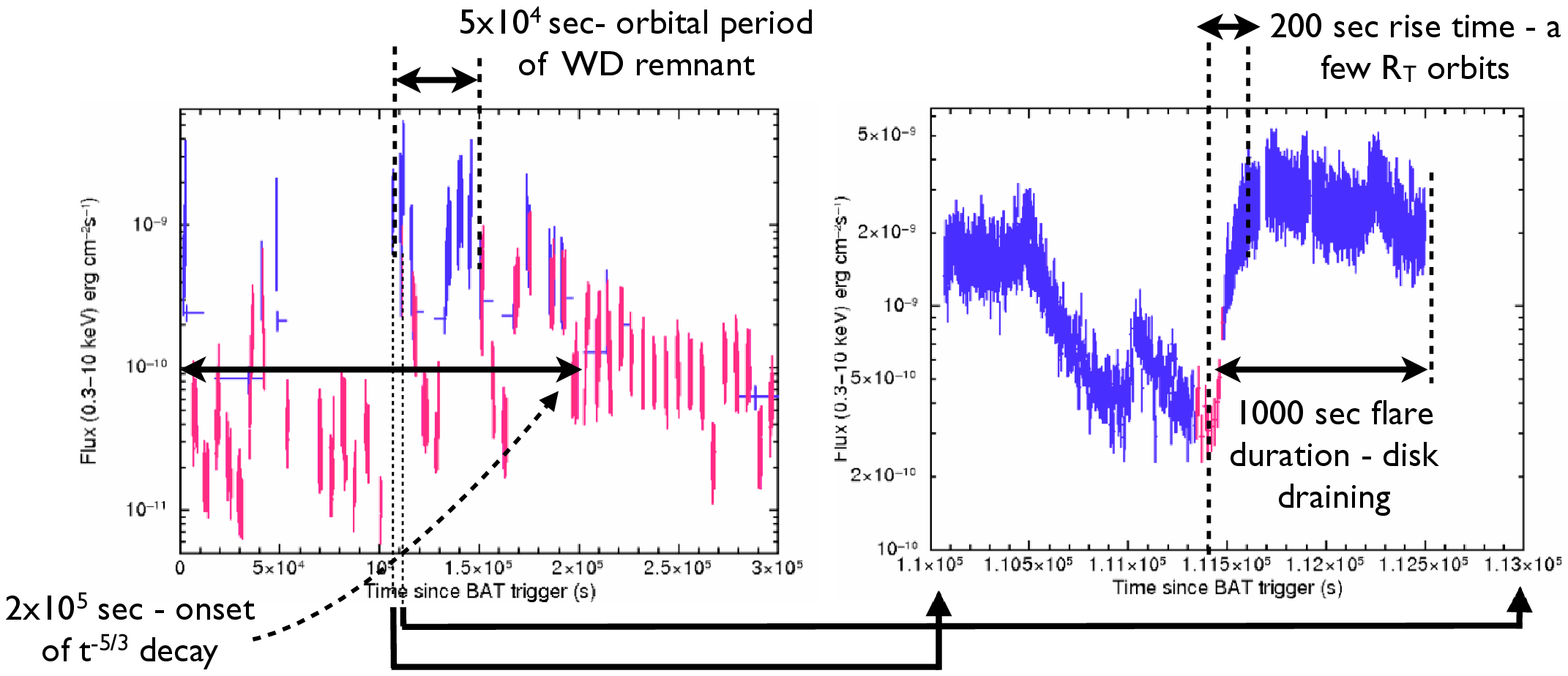}
\vskip -2cm
\caption{Top: A schematic description of our model, a white dwarf in a highly eccentric
orbit that passes near a massive black hole.  A significant fraction of the white dwarf mass is torn
off around the tidal radius, $R_T$. The debris produces a small but massive accretion disk that supports
a jet.  Once the white dwarf moves away from the black hole the tidal disruption ceases, the disk is
drained, and the jet dies out. Bottom: Different observed time scales and their relation to the schematic
model. 
\label{fig:Schematic1}}
\end{figure}

       In this paper, we propose an alternative version of the tidal disruption model
that we believe holds some promise for supplying answers to all of these questions.
We suggest that the star that is tidally disrupted is a white dwarf, not a main
sequence star, and that it is not disrupted all at once, but instead loses pieces
of itself in several passes before dissolving (see Fig. \ref{fig:Schematic1} for a schematic cartoon)
an idea previously considered
in the context of white dwarfs on more nearly circular orbits
by \cite{Sesana+2008} and \cite{Zalamea+2010}.  This suggestion is
motivated by the fact that the fundamental timescale of a tidal disruption is dictated by
the mean density of the star; the greater density of a white dwarf makes it
much easier to achieve the short timescales of this event.  In both our picture and
the main sequence star model, the accretion flow drives a jet, which produces the
observed radiation; as we shall see, this, too, is favored quantitatively by the higher
density of a white dwarf.  The remainder of this paper will develop the consequences of
these ideas.

Note that tidal disruption of a white dwarf by a black hole has been also
previously discussed  in the context of the possible nuclear ignition of
the white dwarf \citep{WilsonMathews04,Dearborn+05,Rosswog+08}. This requires a rather
small black hole and a deep encounter, with pericenter a small fraction of $R_T$. 
Here we focus on larger black holes and more distant tidal disruptions that do not lead
to such an explosion.

\section{Timescales}

Using the
Using the mass-radius relations of \cite{Nauenberg72}, we find that a white dwarf is disrupted at
\begin{equation}
R_T/R_g \simeq 8 M_{max}^{1/3}{\cal M}_*^{-2/3} M_{BH,4}^{-2/3}
            \left[1.  - 0.64 ({\cal  M}_*/M_{max})^{4/3} \right] ,
\end{equation}
{where $M_{max}\simeq 1.4$ is the maximal mass of a white dwarf in units of $M_\odot$.}
In this estimate, we used an apsidal motion constant $k=0.14$
\citep{sirotkin_kim09} and a binding energy factor $f=6/7$, both appropriate to
an $n=3/2$ polytrope.  Note that we have changed our fiducial black hole
mass from $10^6$ to $10^4 M_{\odot}$ because the higher densities of white
dwarfs require smaller mass black holes in order to keep $R_T/R_g > 1$.  In
fact, white dwarf tidal disruption requires
$M_{BH} < 2 \times 10^5 M_{\odot}$,
and somewhat less than that if the black hole rotates slowly or it is desired
that $R_T$ exceed the ISCO.  The period of a circular orbit at the white dwarf
tidal radius is (ignoring black hole spin)
\begin{equation}
P_{\rm orb}(R_T) \simeq 6 {\cal M}_*^{-1}\hbox{~s},
\end{equation}
three orders of magnitude shorter than for a disrupted main sequence star.
In this regard, it is worth noting that the white dwarf mass distribution
appears to be centered at $\simeq 0.6$--$0.7 M_{\odot}$ \citep{hansen_liebert03}.

Matter torn from a white dwarf, like that taken from a main sequence star, will travel
initially on a wide range of highly-eccentric orbits.   Following the arguments
of \cite{rees88}, the most tightly-bound matter might be expected to have an orbital
period $\sim 900 M_{BH,4}^{1/2} {\cal M}_*^{-3/2}$~s.  However, there are several
potential mechanisms that could substantially shorten this period.  If the black hole mass
is not greatly smaller than its upper bound, the orbital pericenters will be in the
strongly relativistic regime.  There, as pointed out by \cite{cannizzo90}, because
relativistic effects cause differential pericenter precession, different stream
orbits can intersect, so that dissipation transfers energy from orbital motion to
heat and possibly photons.  This process may be able to circularize the orbits on a
timescale of a few orbital periods.  {When the black hole spins rapidly and its
mass is close to the upper bound for tidal disruption, $R_T/R_g$ may be small enough that the orbital
precession rate becomes comparable to the orbital frequency, enhancing this effect (we
thank the anonymous referee for this suggestion).}

Multi-pass tidal break-ups may also lead directly to tighter orbits for tidally-disrupted
matter.  {The orbital time of the white dwarf remnant is $\sim 10$ disk drainage times,
so the only mass remaining in the disk by the time the remnant returns is the mass pushed
outward absorbing the angular momentum of the accreted mass.}  If the mass remaining in
the disk from the previous encounter is a
fraction $\epsilon$ of the newly-arriving mass, the momentum of the matter breaking
off the star near $R_T$ will be reduced by that fraction and its kinetic energy reduced
by $2\epsilon$.  The semi-major axis of the resulting orbit for matter with the mean
incoming orbital energy (i.e., near zero) is then $R_T/(4\epsilon)$,
giving an orbital period $P_{\rm orb}(R_T)(4\epsilon)^{-3/2}$.  If $\epsilon$ is as
little as 0.05, the resulting period is only $\sim 10 P_{\rm orb}(R_T)$.  Because
there was a precursor to this event, it is possible that this sort of process
may have occurred as early as the flare that caused the BAT trigger.

As remarked before, internally-generated magnetic field requires $\sim 10$ orbital
periods to reach saturation.  When the period is as short as $P_{\rm orb}(R_T)$ for a
white dwarf tidal disruption, the magnetic field
can grow to full strength within $\sim 100$~s.  Such rapid growth would then be
consistent with explaining the rapid rise time of the flares.
 
The $\sim 1000$--10000~s duration of individual flares might now be reinterpreted as the
drainage time for the small, temporary accretion disk formed from captured material.
Once the turbulence has reached saturation, this timescale
$t_{\rm in} \sim (P_{\rm orb}/2\pi) \alpha^{-1} (R_T/H)^2$ if matter
moves inward only by the action of internal disk stresses, where $\alpha \sim 0.1$ is
the ratio of vertically-integrated magnetic stress to vertically-integrated pressure,
and the disk density scale height is $H$.  {Close to the ISCO, however,
the mean stress overestimates the inflow time because stress fluctuations can
remove enough angular momentum from fluid elements to send them all the way into the
plunging region, permanently removing them from the disk.  As a result,} \cite{khh06} found
that for $R < 3R_{ISCO}$ in a disk with saturated MHD turbulence,
$t_{\rm in} \sim 10 P_{\rm orb}(R)$.  Allowing for some material on orbits with periods
greater than $P_{\rm orb}(R_T)$, $\gtrsim 1000$~s might then be a reasonable
estimate for the total duration of a single flare, including both build-up of the
magnetic field and inflow.  

If the initial pericenter of the white dwarf's orbit is $\sim R_T$, in its first passage
it may lose only a fraction of its mass (indeed the detection of a precursor flare
$\sim 3$~days before the BAT trigger [\cite{burrows11}] suggests that the
very first passage may have been slightly outside $R_T$).
If that is the case, some of its mass is captured into orbiting streams by the
black hole, but the rest of the star remains a coherent bound object, albeit substantially
stretched and distorted.  It can then make several more passes through the region near
$R_T$, traversing a highly eccentric, but well-defined, orbit.

If its initial approach was on a parabolic orbit, the white dwarf's net binding energy to
the black hole is likely to be comparable to its initial self-gravitational binding energy
because the energy to alter its structure is taken from the orbital energy.  The
period of its orbit after capture is then approximately
\begin{equation}
P_{\rm orb}(R_c) \sim {2\pi \over (2f)^{3/2}} \left({M_{BH} \over M_*}\right)
\left({R_*^3 \over GM_*}\right)^{1/2} \sim 6 \times 10^4 {\cal M}_*^{-2} M_{BH,4}
\hbox{~s}.
\end{equation}
It is natural to identify this period with the interval between flares, as each time
the star passes through the pericenter of its orbit it loses a fraction of its mass
(see Fig. \ref{fig:Schematic1}).
Moreover, because in each of these passes additional gravitational work is done stretching
the remnant of the white dwarf, its orbit changes from pericenter passage to pericenter
passage, explaining the irregularity of the inter-flare intervals.  These
structural changes may also alter the effective $R_T$ from one passage to the next.

{When the white dwarf is finally disrupted completely (presumably just before the
final flare at $\sim 1.7 \times 10^5$~s after the initial BAT trigger), its remaining
mass finds itself spread over orbits with a wide range of binding energies and periods.
The tidal streams with orbital periods longer than the remnant's orbital period return
to the neighborhood of $R_T$ only after a comparatively long time, extending the duration
of the event.  If the distribution function of tidal-stream mass with orbital energy is roughly
flat \citep{rees88,phinney89,lkp09}, the late-time accretion rate should decline
$\propto t^{-5/3}$ from the time at which the white dwarf is completely disrupted onward.}

\section{Accretion physics and driving the jet}

In the interior of a white dwarf, the electrons are highly degenerate.
The equivalent temperature of the Fermi level $E_F$ is
$2 \times 10^9 \rho_6^{2/3}$~K, where we scale to a characteristic density of $10^6$~gm~cm$^{-3}$ 
because the mean density of a white dwarf is
$1.1 \times 10^6 {\cal M}^2 M_{max}^{-1}/ (1- 0.64 ({\cal M}/M_{max})^{4/3})$~gm~cm$^{-3}$
(again using the mass-radius relation of \cite{Nauenberg72}).
On the other hand, typical interior temperatures are $\sim 10^7$~K
\citep{hansen04}.  The initial phases of tidal disruption do not change this degree
of degeneracy because adiabatic expansion leaves the ratio $E_F/kT$ invariant \citep{ll80}.

The flow does not remain adiabatic for long, however. For example, orbital
precession leads to shocks.  In these shocks, electron heating will initially
be retarded by scattering-suppression due to their degeneracy, but there is
no such constraint on the ions.  The slowest ion-electron heating rate occurs
when the ions are so much hotter than the electrons that the relative velocity
is dominated by the ion thermal speed.  Before allowing for electron degeneracy,
the characteristic time for heat transfer by ion-electron Coulomb scattering
in the disk formed by tidal disruption of a white dwarf is
\begin{equation}
t_{ion,heat} \sim 8 \times 10^{-6} (H/R) M_{BH,4} T_{i,10}^{3/2} {\cal M}_*^{-3}
     (\Delta M/M_*)^{-1}\hbox{~s}.
\end{equation}
Here we have set the Coulomb logarithm to 30 and assumed that, appropriate to
a C/O composition, the mean ion mass is $14m_p$.  The ratio $\Delta M/M_*$ is
the fraction of the white dwarf mass deposited in a disk with radius $R_T$ and
scale height $H$.  Degeneracy retards energy
transfer to electrons because the crowded phase space partially suppresses collisions.
Only electrons with initial momenta close enough to the Fermi level momentum that
the momentum transfer per event $\Delta p_{ie} \sim m_e \langle v_i\rangle$ can
lift them to the region of unoccupied states can participate.  The fraction of
electrons in the total population able to scatter is then
\begin{equation}
f_{\rm scatter} \sim 3 (m_e/2m_i)^{1/2}(kT_i/E_F)^{1/2} \sim 0.05 (kT_i/E_F)^{1/2}
\end{equation}
if $\Delta p_{ie} \ll p_F$.  The unadjusted heating time is so short that it is
hard to imagine circumstances in which the degeneracy correction could make
any difference on the timescales relevant to this situation.

Thus, the electrons and ions can be expected to thermally equilibrate very rapidly,
and at these densities and temperatures, the electrons (now no longer degenerate) will
also rapidly thermally equilibrate with radiation.  However, the radiative cooling
time of the system is much longer than any of the relevant timescales:
\begin{equation}
t_{\rm cool} \sim 1 \times 10^{11} (H/R) (\Delta M/M_*) {\cal M}_*^{7/3} M_{BH,4}^{-2/3}
\hbox{~s}.
\end{equation}
The radiation pressure is therefore effectively trapped within the material.  Because
shrinking the highly eccentric orbits to nearly circular requires dissipating an
energy comparable to the orbital energy, the disk can therefore be expected to be
geometrically thick, $H/R \sim 1$.
This is the regime of photon-trapping
associated with super-Eddington accretion \cite{begelman79,abramowicz88}.  Because
the diffusion time is long compared to the inflow time, the radiation intensity
distribution is far from steady-state, and the emergent luminosity is much less than
the the rate at which heat is dissipated.  Consequently, the ratio of thermal disk
luminosity to rest-mass accretion rate is much less than the conventional $\sim 0.1 \dot Mc^2$.

The total pressure in the inner disk can be expected to be of order the
electron density times the local virial temperature,
\begin{equation}
p_{\rm disk} \sim 5 \times 10^{21} (\Delta M/M_*) {\cal M}_* M_{BH,4}^{-3} (H/R)^{-1}
        \left({R \over 10R_g}\right)^{-4}\hbox{~dyne~cm$^{-2}$}.
\end{equation}
If the magnetic pressure on the event horizon is limited by the inner disk pressure
(the simulations of \cite{beckwith09} suggest that it may be a factor of 3--4 smaller),
the expected field strength would be
\begin{equation}
B_{\rm hor} \lesssim 4 \times 10^{11}(\Delta M/M_*)^{1/2} {\cal M}_*^{1/2} M_{BH,4}^{-3/2}
         (H/R)^{-1/2}\left({R \over 10R_g}\right)^{-2}\hbox{~G}.
\end{equation}
One might then predict a jet Poynting power as much as
\begin{equation}
L_{\rm jet} \sim 3 \times 10^{50}\phi (B_{\rm hor}^2/8\pi p_{\rm disk})(\Delta M/M_*)
          {\cal M}_* M_{BH,4}^{-1}\hbox{~erg/s}.
\end{equation}
Even after allowing for a field rather less intense than the disk pressure,
$\Delta M/M_* < 1$, and a small coefficient $\phi$, it would seem that the jet power
could easily reach the level seen (peak power after allowance for beaming of
$\sim 3 \times 10^{46}$~erg~s$^{-1}$).  Note, also, that
$L_{\rm jet} \propto M_{BH}^{-1}$, so models requiring larger black holes tend
to generate weaker jets.   As pointed out by our anonymous referee, the smaller
(in $R_g$ terms) disks characteristic of white dwarf disruptions also minimize
Compton drag \citep{phinney87}.

Numerical general relativistic MHD simulations of accretion have shown that
the luminosity of the jet can also be estimated in terms of an effective ``efficiency" per
unit rest-mass accreted that is a function of the black hole spin \citep{mck05,hk06}.
Using the expression found in the latter reference, one might predict
\begin{equation}
L_{\rm jet} \sim 4 \times 10^{49} \left[1 - |a/M|\right]^{-1} (\Delta M/M_*)
           {\cal M}_*^2\hbox{~erg/s}
\end{equation}
if the inflow time is $\simeq 10 P_{\rm orb}(R_T)$.  For a black hole with spin
parameter $a/M \simeq 0.9$, this estimate agrees with the previous one evaluated
for the fiducial parameters.  That it should do so is no coincidence---as
assumed in the previous estimate, the magnetic field intensity on the horizon
is comparable to the inner disk pressure in these simulations.
However, it should be borne in mind that the energy for this jet is actually drawn
from the reducible mass (the rotational kinetic energy) of the black hole, not the
accretion flow.  The function of the accretion is solely to sustain a strong
magnetic field on the black hole horizon.

  Although nominally independent of black hole mass, in fact this second estimate
has an implied dependence through the bound on black hole mass placed by the density
of the disrupted star.  In rough terms, the second jet luminosity estimate
simply mirrors the accretion rate, which is $\propto \Delta M \Omega(R_T)$.  In both
the main sequence star and
white dwarf models, the amount of mass placed in orbit is $\sim M_{\odot}$.
Where they differ is exactly the point we have emphasized in the context of
the lightcurve's timescales: $\Omega(R_T) \sim (G\rho_*)^{1/2}$, which is
$\sim 10^3$ times larger for white dwarfs than for main sequence stars.

When the disk is drained, the inner disk pressure falls.   If that permits the
flux on the horizon to expand, so that the magnetic field there becomes weaker,
the jet would be correspondingly diminished.   Without continuing accretion---that
is, between episodes of tidal capture---radiation from the jet would be much reduced.
This is, of course, what one would expect in those intervals when the remnant of
the white dwarf is out near apocenter.

\section{Summary}

    We propose that the remarkable event known as Swift J1644+57 is more likely
the tidal disruption of a white dwarf than a main sequence star.
{The fact
that white dwarfs are typically $\sim 10^6$ times denser than main sequence
stars makes it much easier to understand the very short timescales characteristic
of this object's lightcurve.  Postulating that the initial encounter was not
close enough to disrupt the white dwarf completely explains the remarkable flares
seen over the event's first few days {\color{red}(see Fig.~\ref{fig:Schematic1})}.
The $\sim 100$~s fluctuations during the
flares may then be identified with the inflow timescale from radii $\sim R_T$, which
in this case is likely only $\sim 10R_g$.  This timescale is imprinted on the
lightcurve if the radiation emerges from the jet when it has traveled less than
a few hundred seconds (in the observer's frame) from the black hole, a distance
equivalent to $\sim 10^3$--$10^4R_g$.  The disk drainage time is longer
because the disk can be expected to spread to radii somewhat larger than $R_T$; this
accounts for the flare durations, $\sim 10^3$--$10^4$~s.  The interflare time
$\sim 5 \times 10^4$ is the orbital period of the white dwarf remnant.  Because
the remnant orbital period is rather longer than the drainage time, there is an
extended period after most of the disk mass has been accreted, but before it is
refilled by the next pericenter passage of the white dwarf, when the accretion
rate is low, and the jet is therefore weak.}

{The higher densities of white dwarfs also explain
both why the observed radiation is dominated by a jet and the jet's high luminosity.
The very high surface density of the disk thoroughly traps thermal photons, while
the associated high pressure can support a stronger magnetic field on the black hole.}
Although the jet luminosity associated with a main
sequence star event is nominally just enough to supply the observed radiated power
(after beaming corrections), that may still be inadequate.  The numbers quoted
refer only to luminosity in observed bands; it is possible there is additional
luminosity elsewhere in the spectrum (e.g., between 100~keV and 100~MeV).
More importantly, in most jet radiation models (e.g., \cite{celottighisellini08}),
the photon luminosity is only $\sim 1\%$ -- $10\%$ of the
kinetic power; the jet powers estimated on the basis of the disk
pressure or accretion rate refer to the total, so that $L_{\rm jet}$ must be
significantly {\it greater} than the observed luminosity.

     Although the space density of white dwarfs ($\simeq 3 \times 10^{-3}$~pc$^{-3}$
in the Solar neighborhood: \cite{rowell_hambly11}) is very similar to the space density
of solar-mass main sequence stars ($\simeq 3.5 \times 10^{-3}$~pc$^{-3}$:
\cite{reid02}), their effective cross section for coming close
enough to a black hole to be tidally disrupted is smaller by the ratio of their tidal
disruption radii,
$\simeq 2 \times 10^{-3} ({\cal M}_{\rm *,wd} {\cal M}_{\rm *,ms})^{-2/3}$.
\cite{burrows11} estimated the total number of solar-mass main sequence
star disruptions within the volume detectable by Swift to be $\sim 10^4$~yr$^{-1}$,
suggesting that the rate of white dwarf tidal disruptions should be $\sim 10$~yr$^{-1}$.
{\color{blue}If they are all relativistically beamed so that we see only $\sim 1\%$ of
all events, the observable rate falls to $\sim 0.1$~yr$^{-1}$, and is reduced further
by a factor $\sim 10$ to account for BAT's field of view and observing efficiency
\citep{burrows11}.  Thus, we might expect BAT to detect perhaps $\sim 0.01$~yr$^{-1}$,
rather than the actual $\sim 1/6$~yr$^{-1}$.  On the other hand, the rates estimated
by \cite{burrows11} (and similarly by \cite{zauderer11}) were based on theoretical
predictions that are an order of magnitude smaller than some empirically-based rate
estimates \citep{Gezari+08,maksym10}.  Given the many uncertainties (and small-number
statistics), we believe the rate of white dwarf disruptions is consistent with the
Swift detection rate.}  Moreover,
a large part of the contrast between our
predicted rate and the rate expected for main sequence stars would be
removed if the required pericenter distance for a main sequence tidal disruption is
$\sim 0.1 R_T$, as suggested by \cite{cannizzo11}.

    We close with a final significant contrast between white dwarf and main
sequence star models for this event.  The black holes most effective at tidal
disruption of white dwarfs are smaller by the ratio
$(\rho_{\rm *,ms}/\rho_{*,wd})^{1/3} \sim 10^{-2}$.  Although there has
long been good evidence for black holes of $\sim 10^6 M_{\odot}$ or more in the central regions
of Galaxies (beginning with the Milky Way) and somewhat shakier evidence for black
holes in galactic nuclei with masses $\gtrsim 10^5 M_{\odot}$ (collected in \cite{xiao11},
these masses are all based on the assumption that gravitational dynamics dominate
broad-line gas motions, and in most cases use only scaling arguments to estimate
the broad-line region's size),
this event represents the first indication of a black hole in the
$\sim 10^4$--$10^5 M_{\odot}$ mass range.  On the basis of
the $M_{BH}$--bulge luminosity correlation, \cite{burrows11} estimated that this
galaxy should have a nuclear black hole with mass $\simeq 2 \times 10^7M_{\odot}$.
If we are correct in our suggestion that this event was due to disruption of
a white dwarf rather than a main sequence star, the correlation estimate may be too
large by a factor of 100 or more in this galaxy.  Alternatively, the black hole responsible
for this disruption could be a second, smaller black hole orbiting a larger one, whose
mass is closer to the correlation estimate.

\acknowledgements{We thank Phil Evans for supplying the Swift light curves, Re'em Sari for
helpful discussions and an anonymous referee for several insightful suggestions.  This work was
partially supported by NSF grants AST-0507455 and
AST-0908336 (JHK) and by an ERC advanced research grant and the ISF center for High
Energy Astrophsics (TP).  One of us (JHK) would also like to thank the Racah Institute
for Physics at the Hebrew University for its generous hospitality while much of
the work on this paper was accomplished.}



\end{document}